\documentclass[entropy,article,accept,pdftex,moreauthors]{mdpi} 

\firstpage{1} 
\pubvolume{25}
\issuenum{8}
\articlenumber{1114}
\pubyear{2023}
\copyrightyear{2023}
\externaleditor{Academic Editor: Santi Prestipino}
\datereceived{ 26 June 2023} 
\daterevised{21 July 2023 } 
\dateaccepted{ 21 July 2023} 
\datepublished{ 26 July 2023} 
\hreflink{https://doi.org/10.3390/e25081114} 
\pdfoutput=1 
\Title{The Solid Phase of $^4$He: A Monte Carlo Simulation Study}

\TitleCitation{The Solid Phase of $^4$He: A Monte Carlo Simulation Study}

\Author{Massimo Boninsegni \orcidA{}}

\AuthorNames{Massimo Boninsegni}

\AuthorCitation{Boninsegni, M.}

\address[1]{
\quad Department of Physics, University of Alberta, Edmonton, Alberta, Canada T6G 2H5}

\corres{Correspondence: m.boninsegni@ualberta.ca}

\abstract{The thermodynamics  of solid ({hcp})  $^4$He is studied theoretically by means of unbiased Monte Carlo simulations at finite temperature, in a wide range of density. 
This study complements and extends previous theoretical work, mainly by obtaining results at significantly lower temperatures (down to 60 mK) and  for systems of greater size, by including in full the effect of quantum statistics, and by comparing estimates yielded by different pair potentials. 
All the main thermodynamic properties of the crystal, e.g., the kinetic energy per atom, are predicted to be essentially independent of temperature below $\sim$ 1 K. Quantum-mechanical exchanges are virtually non-existent in this system, even at the lowest temperature considered. However, effects of quantum statistics are detectable in the momentum distribution.
Comparison with available measurements shows general agreement within the experimental uncertainties.}


\begin{document}


\section{Introduction}
\label{introd}
Among naturally occurring substances, the solid phase of helium (which is stable only under applied pressure) represents a unique example of ``quantum crystal'', i.e., one in which quantum mechanics most considerably and observably affects its  thermodynamic properties \cite{Wilks1967}. 
The most direct manifestations include a kinetic energy per particle markedly above its classical value   3$T$/2 (we adopt a system of units in which the Boltzmann constant $k_B=1$); quantitative theoretical predictions for such excess kinetic energy have been made \cite{cuccoli1993} for all rare gas solids, and there exist a number of experimental confirmations (see, for instance, Ref. \cite{Timms1996}). For the rare gas solids of elements heavier than helium, however, the excess kinetic energy is essentially the {    only} measurable manifestation of their quantum-mechanical nature. Other relevant physical properties (e.g., the momentum distribution) can be largely understood along classical lines.
\\ \indent 
In solid helium, on the other hand, the classical picture is quantitatively altered by zero-point motion, which is nowhere near as important in other solids (with the sole exception, albeit not to the same extent, of molecular hydrogen \cite{Dusseault2017}. This includes enhanced atomic excursions away from the equilibrium (lattice) positions, with respect to what is predicted and observed in most solids, based on classical (thermal) arguments. Indeed, for a long time, zero point motion was believed to be the physical reason underlying the failure of liquid helium to solidify, under the effect of its own vapor, a view that has  relatively recently been challenged, as quantum statistics (i.e., exchanges of indistinguishable particles) has been shown to play a pivotal role in the stabilization of the crystal phase, at least in $^4$He \cite{Boninsegni2012a}. In solid $^3$He, exchanges have been proposed as the physical agent responsible for the stabilization of the {    bcc} over the {    hcp} crystalline phase \cite{Pederiva1998}.
\\ \indent
As helium can be regarded as the archetypal quantum crystal, it poses a challenge to many-body theorists; achieving an accurate microscopic description, based on realistic interatomic potentials, capable of making reliable predictions for experimentally cogent quantities, is a worthwhile goal in theoretical condensed matter and many-body physics. Of particular interest are calculations in which the fundamental quantum-mechanical equations are treated  without any uncontrolled approximations. Ideally, one starts from a microscopic Hamiltonian in which the only external (independent) input is the potential energy of interaction among atoms.
\\ \indent
Experimentally, the bulk of the available information on the solid phase of helium has been delivered by x-ray \cite{Fain1970,Heald1983,Mao1988,Loubeyre1993} and neutron \cite{Henshaw1958,Werthamer1970,Svensson1986,Simmons1987,Diallo2004,Adams2007,Diallo2007} scattering measurements, whose main outcome is the thermodynamic equation of state, as well as single-atom dynamics through the momentum distribution, from which the mean atomic kinetic energy can be inferred \cite{Azuah1997}.
On the theoretical side, Path Integral Monte Carlo simulations have played a major role in shaping our current understanding of the physics of condensed helium in the condensed phase \cite{Ceperley1995,Herrero2014}. These are unbiased calculations which involve no {    a priori} physical assumptions. Indeed, only two main assumptions are built into the (almost) totality of PIMC simulations, namely (a) that He atoms, which are taken as the elementary constituents (see discussion below) can be regarded as {    distinguishable} particles, as quantum-mechanical exchanges are suppressed in the solid phase, due to atomic localization and (b) that the bulk of the potential energy of interaction among the helium atoms can be captured by a pair-wise, central potential featuring a repulsive core at short interatomic separation; three-body interaction terms are neglected.
\\ \indent 
The development of an accurate, {    ab initio} pair potential, describing the interaction of two helium atoms, has been the subject of an intense theoretical effort spanning nearly a century and still ongoing \cite{Mason1954,Aziz1979,Ceperley1986,Przybytek2017}. The first pair interaction to have been extensively used in PIMC simulations whose results have been compared to available experimental data for the superfluid phase of $^4$He \cite{Ceperley1986b} is the earliest version of the Aziz pair potential \cite{Aziz1979}. The agreement between theory and experiment afforded by that potential is generally satisfactory, as confirmed by subsequent studies \cite{Boninsegni2006,Prisk2017} 
overcoming the main limitations of the original calculation, mainly the relatively small size of the simulated system (64 atoms).
\\ \indent
For the solid phase, on the other hand, the comparison of experimental and theoretical estimates has been considerably more limited in scope, for a number of reasons. The most important is the simple fact that the wealth of experimental data available for the superfluid phase \cite{Barenghi1998} is presently unmatched in the crystal. The resurgence of interest in the solid phase of helium that took place two decades ago, following the report of possible ``supersolid'' behavior \cite{kim2004}, prompted new investigations, significantly expanding the available comparison dataset.
However, assessing different theoretical calculations is complicated by the plethora of helium pair potentials \cite{Aziz1987,Montgomery1989,Aziz1992,Aziz1995,Janzen1997, Korona1997,Przybytek2010,Przybytek2017} that have been proposed and utilized since Ceperley and Pollock's 1986 calculation. 
\\ \indent 
Many of these potentials are either refinements of, or at any rate based upon the original Aziz model interaction; the differences among them are often relatively small and pertain to specific features, such as the depth of the attractive well, or the behavior of the repulsive core at short interatomic separation. To be sure, some of these aspects become quantitatively important as the system is compressed, and for pressures, e.g., in the megabar range, three-body interactions must be included in order to achieve a quantitative reproduction of the experimental EOS \cite{Boninsegni1993,Chang2001}. 
It is not clear, however, how sensitive to the details of the pair interaction are the most important experimental quantities, e.g., the single-particle kinetic energy, especially at low pressure (less than $\sim$ 100 bars). 
\\ \indent
We present here the results of PIMC simulations of the {    hcp} phase of $^4$He at low temperature (typically around 1 K, but also as low as $\lesssim 0.1$ K for specific densities), in a range of pressure between 25 and 150 bars. We consider a perfect helium crystal, namely one free of point (e.g., vacancies or interstitials) or extended (e.g., dislocations) defects. We make use of different versions of the Aziz pair potential and compare the results for  the single-particle kinetic and total energy per atom and for the pressure to the most recent experimental estimates. For the physical quantities considered here, agreement with available experimental data seems satisfactory; the differences between the estimates yielded by the different potentials are typically (much) smaller than the experimental uncertainty. In other words, in the region of parameter space considered here, all of the pair potentials that have been utilized over the past two decades give essentially equivalent results.
This study also provides additional, strong evidence that a perfect crystal of $^4$He is {    not} superfluid, at arbitrarily low temperatures. Indeed, quantum-mechanical exchanges of atoms remain strongly suppressed even at a temperature at low as 60 mK, at the melting density. Nevertheless, the effects of quantum statistics are detectable in the one-body density matrix (and therefore, experimentally, in the momentum distribution). We also compare our low-temperature energy estimates to recent ground-state Monte Carlo calculations; the comparison again points to the greater reliability of finite temperature techniques.
\\ \indent
This paper is organized as follows: in section \ref{model} we describe the model of interest and briefly review the computational technique utilized; in section \ref{res} we present our results, offering a discussion in section \ref{concl}.

\section{Model and Methodology}
\label{model}
The system is described as an ensemble of $N$ point-like, identical particles with a mass $m$ equal to that of a $^4$He atom, and with spin zero, thus  obeying Bose statistics. This is where a basic assumption is built into the model, namely the Born-Oppenheimer approximation, which allows us to integrate out the electrons and regard $^4$He atoms as elementary particles. At ordinary conditions of temperature and pressure, this procedure is justified by the large decoupling between electronic and ionic energy scales. In different physical settings, e.g., in the interior of Jovian planets, such separation is no longer possible, and one has to take into account electronic and ionic degrees of freedom separately \cite{Morales2009}, but, in this study, we regard the individual $^4$He as our elementary constituents.

The system is enclosed in a cell of volume $\Omega$ so that $n=N/\Omega$ is the nominal density. The cell is shaped like a cuboid, with periodic boundary conditions in all three directions. The sizes are adjusted to fit a perfect classical crystal, with no defects.
\\ \indent
The quantum-mechanical many-body Hamiltonian reads as follows:
\begin{eqnarray}\label{u}
\hat H = - \lambda \sum_{i}\nabla^2_{i}+\sum_{i<j}v(r_{ij})
\end{eqnarray}
where the first (second) sum runs over all particles (pairs of particles), $\lambda\equiv\hbar^2/2m=6.0596415$ K\AA$^{2}$, $r_{ij}\equiv |{\bf r}_i-{\bf r}_j|$ and $v(r)$ is the pair potential which describes the interaction between two helium atoms. As mentioned in the Introduction, several different forms for $v$ have been used in this study. Most of the results shown here were obtained with the original (1979) Aziz pair potential \cite{Aziz1979} (henceforth referred to as Aziz I), which is largely phenomenological, the result of a careful combination of all available theoretical and experimental data for the interaction of two helium atoms in the gas phase. The Aziz I pair potential has been shown to provide a rather accurate quantitative description of the superfluid phase of $^4$He \cite{Ceperley1995}. Generally speaking, at high density (in both the liquid and solid phases), the pressure obtained from calculations based on the Aziz I potential is higher than the experimentally measured one, signaling that the Aziz I potential is too ``stiff'' at short distances and/or the presence of attractive forces, presumably arising from the interaction of clusters of atoms (e.g., triplets).
\\ \indent
Two different strategies have been pursued, in order to improve the agreement between theory and experiment. In the effective potential approach \cite{Young1981}, one attempts to incorporate three-body effects into a single pair potential, which is parameterized to reproduce the experimental equation of state. Alternatively, and this is the approach that has been most extensively considered in recent times, the aim is that of moving away from empirical, parametrized forms, utilizing instead expressions arising from  {    ab initio} quantum chemistry calculations. In that case, three-body terms are included as separate contributions, and this is the spirit in which the original Aziz potential has been subsequently refined and other versions proposed. The result has been typically that of slightly {    worsening} the agreement between theory and experiment, which would then be improved by the explicit inclusion of three-body terms \cite{Cencek2007,Sese2020}.
As mentioned above, however, no systematic comparison has been carried out of different versions of the Aziz pair potential when it comes to reproducing the physical properties of the solid phase of helium at low pressure, where the effect of three-body interactions is expected to be small. Here, for comparison purposes we are going to show results obtained with the Aziz I potential, as well as with the versions thereof proposed in 1987 \cite{Aziz1987} (Aziz II) and 1995 \cite{Aziz1995} (Aziz III).
\\ \indent 
We performed QMC simulations of  the system described by Eq.  (\ref{u}), based on the canonical version \cite{Mezzacapo2006,Mezzacapo2007} of the continuous-space Worm Algorithm \cite{Boninsegni2006} based on Feynman's space-time approach to quantum statistical mechanics \cite{Feynman1965}.  Since this methodology is extensively described in the literature, it will not be reviewed here; rather, all the relevant technical details will be provided, enabling others to attempt to reproduce the results presented here.
The first aspect to discuss is that in all the simulations carried out in this study, $^4$He atoms are treated as {    indistinguishable}, i.e., the many-particle paths in imaginary time are allowed to ``entangle'', with particles trading places \cite{Ceperley1995,Feynman1965}. Previous attempts to assess quantitatively, by computer simulation, the importance of quantum statistics in a perfect crystal of helium concluded that the effect was negligible \cite{Clark2006,Boninsegni2006b}; in this work, we extend those studies to lower temperatures. We note that the WA has been shown particularly effective at sampling  permutations of identical particles, and therefore we may be reasonably confident that failure to detect a significant presence of exchanges should constitute a true physical reflection of the nature of the system, as opposed to possible algorithmic inefficiency.
\begin{figure}[H]
\centering
\includegraphics[width=9 cm]{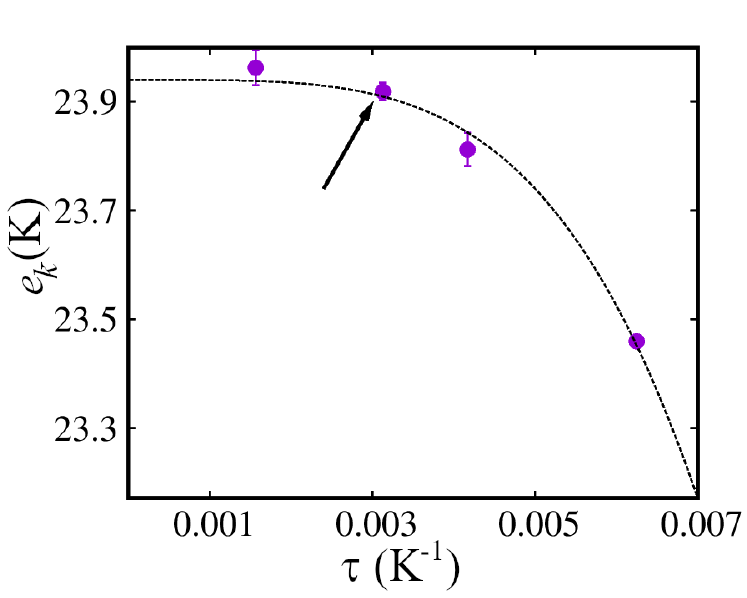}
\caption{Time step extrapolation of the kinetic energy per particle $e_k(K)$ in {    hcp} $^4$He at temperature $T=1$ K, for a system comprising $N=216$ atoms. The density is $n=0.0286$ \AA$^{-3}$. The results are obtained with the Aziz I pair potential.
The solid line is  a quartic fit to the data. Arrow points to the value of $\tau$ for which convergence is observed. \label{fig1}}
\end{figure}   
Details of the simulation are  standard; the short-time approximation to the imaginary-time propagator used here is accurate to fourth order in the time step $\tau$ (see, for instance, Ref. \cite{Boninsegni2005}). We have carried out numerical extrapolation of the estimates to the $\tau\to 0$ limit, and observed convergence of the thermal averages for a value of $\tau=3.125\times 10^{-3}$ K$^{-1}$ for all quantities of interest here. Fig. \ref{fig1} shows the extrapolation procedure for the kinetic energy per atom, $e_k$. Standard estimators were used for all the energetic properties, including the well-known virial estimator for the pressure \cite{Ceperley1995}. 
\\ \indent
Physical quantities of interest calculated in this study include, besides pressure and energetics, structural correlations such as the pair correlation function, which is related to the experimentally accessible static structure factor.  We have carried out simulations for systems of different  sizes, the largest comprising $N=512$ atoms. In general, we found that the numerical estimates of $e_K$ obtained on a simulated system comprising $N=216$ atoms are indistinguishable, within statistical uncertainties, from those obtained on systems of larger sizes. The range of system sizes considered in this work gives us reasonable confidence in our ability to gauge finite-size effects, i.e., that the numerical values quoted herein are representative of the thermodynamic limit, within their statistical uncertainties.
\\ \indent
For the calculation of the (potential) energy per particle and the pressure, we estimated the contribution of particles outside the main simulation cell by assuming a value of the pair correlation function $g(r)=1$, for $r$ greater than half the (shortest) cell side. Based on the observed quantitative consistency (within statistical uncertainties) of results for systems of different sizes, we contend that this procedure is numerically reliable. We come back to this point more quantitatively when discussing our results.

\section{Results}\label{res}
\subsection{Effect of temperature}
\begin{figure}[H]
\includegraphics[width=13.8 cm]{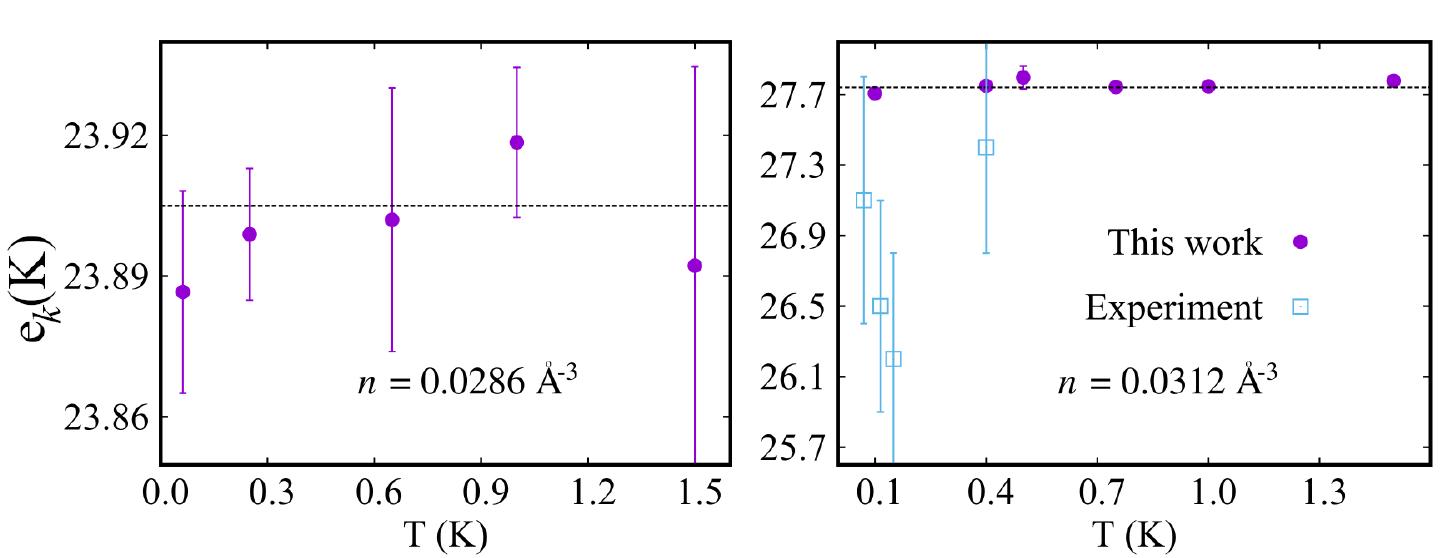}
\caption{Kinetic energy per $^4$He atom $e_k$ computed as a function of temperature, for a {    hcp} crystal of density 0.0286 (0.0312) \AA$^{-3}$, left (right) panel. The simulated system comprises $N=216$ atoms, and the interatomic potential utilized is the Aziz I (1979). Dashed lines represent fits to the data, based on a constant value. Experimental data shown on the right panel are from Ref. \cite{Adams2007}. Not shown in the left panel is the experimental result from Ref. \cite{Diallo2004} at $T=1.6$ K and $n=0.0288$ \AA$^{-3}$, namely 24.25(30) K.}\label{kinetic}
\end{figure}

We begin by discussing the effect of temperature on the energetics and structure of solid $^4$He. All of the calculations whose results are illustrated in this subsection were carried out using the Aziz I pair potential (1979).\\ \indent
The left (right) panel of Fig. \ref{kinetic} shows the computed kinetic energy per $^4$He atom in a {    hcp} $^4$He crystal of density 0.0286 (0.0312) \AA$^{-3}$, in the temperature range $T < 1.5$ K, the lowest temperature being $T\lesssim 0.1$ K. The higher value of density corresponds to a pressure close to 56 bar; the  lower value is very close to the $T=0$ melting density and yields a computed pressure of approximately 25 bars (in both cases the pressure is nearly temperature-independent). Also shown for comparison are the most recent experimental estimates of the kinetic energy per atom in solid helium, namely those of Ref. \cite{Adams2007}.
The two experimental works with which a comparison seems meaningful are Refs. \cite{Diallo2004} and \cite{Adams2007}. The comparison of theoretical and experimental estimates shows quantitative agreement, taking into account the relatively large uncertainty in the experimental determination of the kinetic energy, as well as the uncertainty on the quoted values of the density in Ref. \cite{Adams2007} and the slight differences in temperature. 
\\ \indent
Within the statistical uncertainties of our calculation, no significant dependence on the temperature of the kinetic energy per atom can be detected in the temperature range considered. This is consistent with the observation made in previous theoretical \cite{Ceperley1996} and experimental \cite{Adams2007} work (see Fig. \ref{kinetic}). We generally find this to be the case for all the basic energetic properties, namely kinetic and total energy per atom, as well as the pressure and for structural correlation as well, in the entire range of density considered in this work. Henceforth, therefore, unless otherwise stated, all of the results quoted here are at temperature $T=1$ K, which can be for all practical purposes considered equivalent to the ground state for the helium crystal. We come back to a more comprehensive and quantitative discussion of our results for the energetics of the helium crystal in \ref{energyresults}.
\begin{figure}[H]\centering
\includegraphics[width=9 cm]{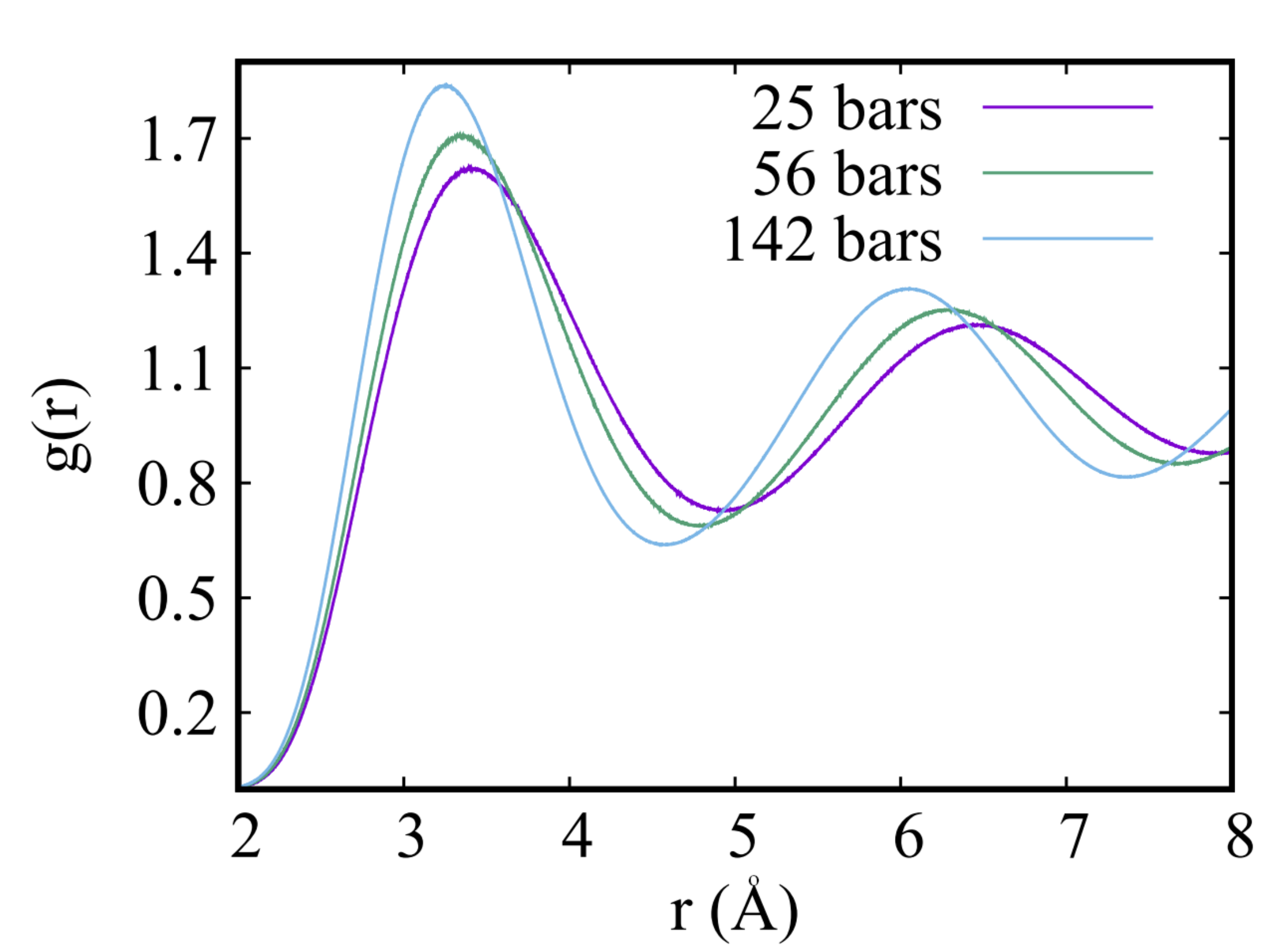}
\caption{Spherically averaged pair correlation function $g(r)$ in {    hcp} solid $^4$He at three different values of the density, at $T=1$ K. The calculations whose results are shown here were performed on a system comprising $N=216$ atoms.}
\label{gofr}
\end{figure}   
Fig. \ref{gofr} displays the spherically averaged pair correlation function $g(r)$, computed by numerical simulations at temperature $T=1$ K, for three different densities, namely $n=0.0286$ \AA$^{-3}$, which corresponds to a pressure of approximately 25 bars, $n=0.0312$ \AA$^{-3}$, for which the computed pressure is 56 bars, and the highest density considered here, namely $n=0.0353$ \AA$^{-3}$, for which the pressure is 141 bars. These are results obtained using the Aziz I potential, but as we show more quantitatively below, the differences in the values of the pressure computed with different versions of the Aziz potential are relatively small (of the order of a fraction of a bar), in the range of densities considered here.  These values of the pressure (and all others reported in Table \ref{tab1} below) are in excellent quantitative agreement with the experimental equation of state of Driessen, van der Polls and Silvera \cite{Driessen1986,note1}.
\\ \indent
The calculations whose results are shown in Fig. \ref{gofr} were performed on a system comprising $N=216$ atoms. As shown in Fig. \ref{gofr}, the value of the pair correlation function $g(r)$ is very close to unity, giving us confidence that the approximation utilized to compute the contribution to the potential energy and to the pressure, namely assuming $g(r)=1$ outside the simulation cell, is quantitatively reliable. One might think of such a procedure as being somewhat ``crude'', but its use is justified, in practice, by both the exponential decay of the fluctuations around unity of $g(r)$, as well as the rapid ($\sim 1/r^6$) decay with distance of the pair potential. In order to obtain a numerical check, we have  carried out calculations on a system of $N=512$  atoms. and computed the pressure based on the same approximation. The results are all consistent with those obtained for a system of $N=216$ atoms; for example, at a density $n=0.0312$ \AA$^{-3}$ we obtain a pressure of 56.3(5) bars at $T=1$ K, on a system of $N=512$ atoms.  
\begin{figure}[H]\centering
\includegraphics[width=8 cm]{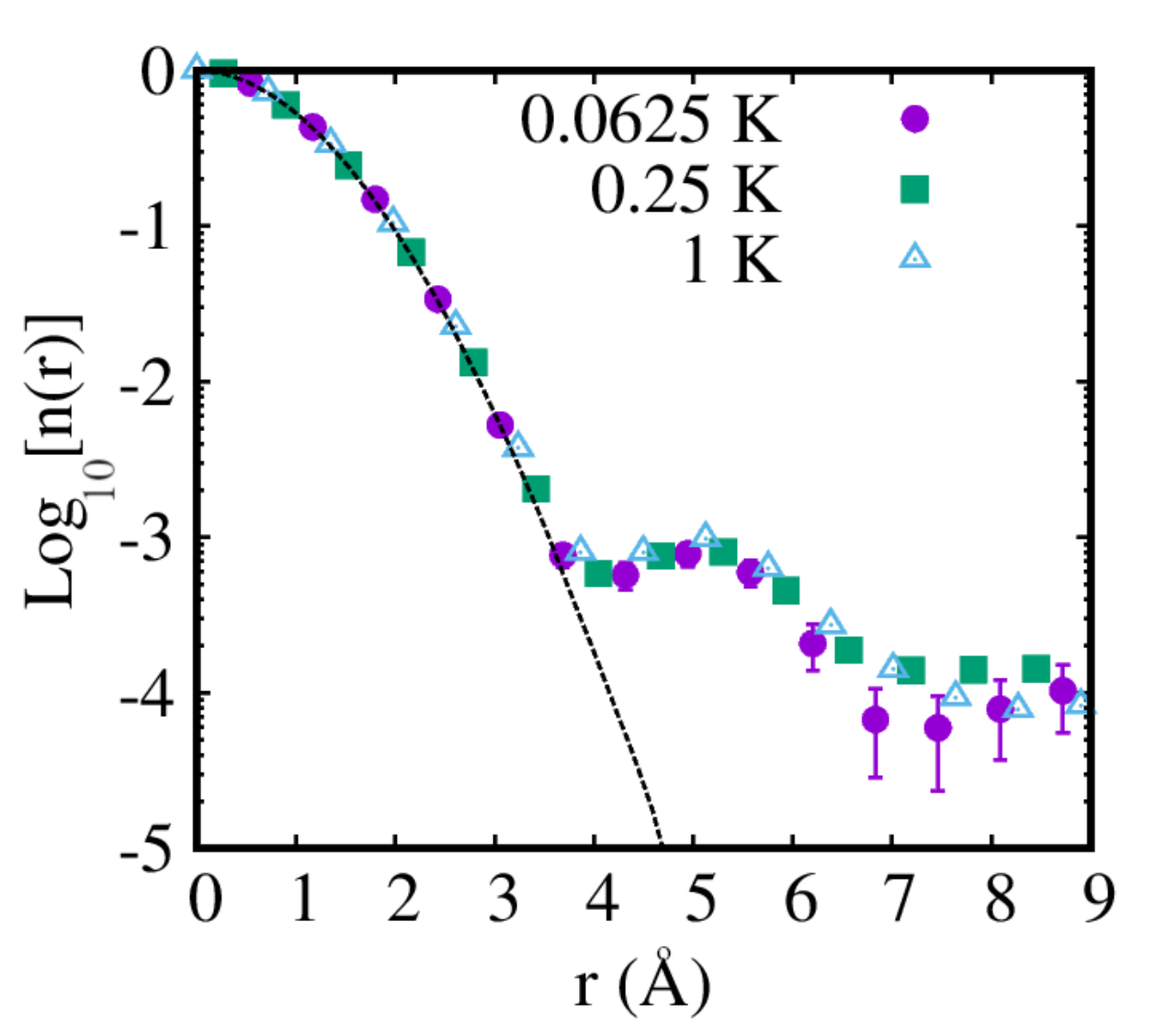}
\caption{Spherically averaged one-body density matrix $n(r)$ in {    hcp} solid $^4$He at three different temperatures. The density is $n$=0.0286 \AA$^{-3}$, and the simulated system comprises $N=216$ atoms. Dotted  line represents the $T=1$ K estimate for a $^4$He crystal in which atoms are regarded as distinguishable ({    boltzmannons}).}
\label{nofr}
\end{figure}   
\subsection{Effects of quantum statistics}
As mentioned in the introduction, the calculations carried out in this work fully include the effect of quantum statistics, which means that exchanges of indistinguishable helium atoms are allowed. However, the frequency with which they occur remains exceedingly low, even at the lowest temperature and density considered here. For example, at $n=0.0286$ \AA$^{-3}$, which is very close to the $T=0$ melting density, and at a temperature as low as $T=0.0625$ K, the measured likelihood of a single $^4$He atom to be involved in an exchange is less than one percent, three-particle exchanges occurring in the basal plane being overwhelmingly the most frequent. Within the statistical errors of our calculation, the effects of Bose statistics on the energetic properties of the helium solid are essentially not detectable. We carried out a separate simulation in which $^4$He atoms are treated as {    boltzmannons}, i.e., distinguishable quantum particles, and estimate the kinetic energy per atom to remain unchanged, within our statistical uncertainties, as a result of the neglect of quantum statistics.
\\ \indent
That is not to say, however, that quantitative effects of quantum statistics cannot be observed, chiefly in the momentum distribution, which is experimentally measurable \cite{Diallo2004,Diallo2007}. Specifically, effects of Bose statistics have been observed in solid $^4$He near melting, at a temperature as high as 1.6 K, through an enhancement of the momentum distribution at low momenta, compared to what one would predict based on a theoretical approach neglecting quantum statistics.
\\ \indent
Our calculations confirm such predictions. Fig. \ref{nofr} displays the spherically averaged one-particle density matrix $n(r)$, which is defined as
\begin{equation}
n(r)=\frac{1}{4\pi \Omega} \int d\alpha \ d^3r^\prime \ \langle\hat\Psi^\dagger({\bf r+r}^\prime)\ \hat\Psi({\bf r}^\prime)\rangle
\end{equation}
where $\langle ... \rangle$ stands for thermal average, $\hat\Psi$ and $\hat\Psi^\dagger$ are the usual Bose field operators and the integration $d\alpha$ is over the solid angle of ${\bf r}$. The one-body density matrix 
is connected to the momentum distribution through a Fourier transformation. The comparison of the one-body density matrix computed with the full inclusion of quantum statistics with that which arises from an identical calculation in which helium atoms are considered distinguishable (dotted line in Fig. \ref{nofr}) shows how the former extends to considerably longer distances, roughly up to three lattice constants. This can be attributed to the increasing particle delocalization associated with quantum-mechanical exchanges. and is consistent with the observation that three-particle ring exchanges are the most frequent. The enhancement of the $n(r)$ at large distances corresponds to a strengthening of the momentum distribution at low momenta.
It is worth noting that a similar manifestation of Bose statistics can be observed in the momentum distribution of other Bose fluids and crystals, e.g., in liquid parahydrogen near freezing \cite{Boninsegni2009}.
\\ \indent
The results shown in Fig. \ref {nofr} also support the conclusions reached in early Monte Carlo studies \cite{Boninsegni2006b,Clark2006}, namely that in a perfect crystal of $^4$He the one-body density matrix decays exponentially at large distances, displaying no noticeable dependence on the temperature, i.e., there is no Bose-Einstein condensation in the $T\to 0$ limit \cite{note2}. Indeed, all credible scenarios of possible ``supertransport'' in the solid phase of $^4$He involve extended defects such as dislocations \cite{Boninsegni2007, Soyler2009,Kuklov2022}; there is overwhelming evidence that  supersolidity is not underlain by point defects such as vacancies or interstistials \cite{Boninsegni2006c,Boninsegni2012b}. Otherwise, theoretical work carried out mainly over the past two decades indicates that the existence of a supersolid phase requires that the pairwise interaction among particles not feature the kind of repulsive core that characterizes instead helium and other rare gas solids \cite{Boninsegni2012}.

\subsection{Energetics}\label{energyresults}
All of our results for the energetics of solid $^4$He are compiled in Table \ref{tab1}. We provides estimates for the total energy per atom $e$ (in K), as well as for the kinetic energy $e_k$ (also in K) and for the pressure, obtained with different versions of the Aziz pair potential, specifically the Aziz I, Aziz II and Aziz III.
\begin{table}
\caption{Energetics of solid $^4$He computed at $T=1$ K by Monte Carlo simulations based on the continuous-space WA. Shown are the total energy per particle $e$ (in K), as well as the kinetic energy $e_k$ (in K), and the pressure in bars. The results are mainly for the Aziz I \cite{Aziz1979} pair potential, but for comparison purposes, some results are included of calculations based on the Aziz II \cite{Aziz1987} and Aziz III \cite{Aziz1995} pair potentials. The results presented in this table pertain to calculations for a system comprising $N=216$ atoms. Statistical errors (in parentheses) are on the last digit. They are not explicitly shown for the pressure, as they are consistently of the order of half a bar. Shown within square brackets are the ground state estimates from Ref. \cite{Rugeles2017}\label{tab1}}
\begin{tabularx}{\textwidth}{CCCCC}
\toprule
\textbf{\textit{n} (\AA$^{-3}$)}	& $e$ (K)	& $e_k$ (K) & Pressure (bars) &Potential\\
\midrule
0.0286		& $-6.00(2)$			& 23.96(2) &24.9 & Aziz I\\
0.0286		& $-6.31(2)$			& 23.93(2) &24.0 & Aziz II\\
0.0286		& $-6.21(2)$			& 24.07(2) &24.1 & Aziz III\\
 & & & & \\
0.0294		& $-5.94(3)$			& 25.11(2) &33.4 &Aziz I \\
0.0294		& $-6.02(2)$			& 25.22(2) &33.1 &Aziz III \\
 & & & & \\
0.0301 & $-5.64(2)$ &26.06(2) &41.7 & Aziz I \\
0.0301 & $-5.79(2)$ &26.30(2) &40.7 & Aziz III \\
 & [$-5.73(1)$] &[26.40(7)] &  & Aziz III \\

 & & & & \\
0.0312 & $-5.20(2)$ &27.75(1) &57.0 & Aziz I \\
0.0312 & $-5.52(2)$ &27.75(2) &55.1 & Aziz II \\
0.0312 & $-5.41(2)$ &27.92(2) &56.7 & Aziz III \\
 & & & & \\
0.0329 & $-4.48(3)$ &30.46(3) &85.6 & Aziz I \\
0.0329 & $-4.59(2)$ &30.62(1) &85.6 & Aziz III \\
 & [$-4.50(1)$] &[30.82(9)] & & Aziz III \\

 & & & & \\
0.0353 & $-2.74(2)$ &34.41(2) &140.7 & Aziz I \\
0.0353 & $-2.89(2)$ &34.73(2) &141.5 & Aziz III \\
& [$-2.83(1)$] &[34.79(8)] &  & Aziz III \\
\bottomrule
\end{tabularx}
\end{table}
The first general remark, regarding the dependence of results on the pair potential utilized in the calculation, is that the differences are very small, much smaller than the current experimental uncertainties. In general, newer versions of the Aziz pair potential yield an energy per atom some $\sim 0.2$ K lower than that afforded by the original (Aziz I) potential. On the other hand, the pressure estimates are very close, whereas those for the kinetic energy are virtually identical for the Aziz I and Aziz II pair potentials, while the Aziz III yields values that are between 0.1 and $ 0.2$ K higher. There is otherwise no detectable difference between the results obtained with the various versions of the Aziz potential for quantities such as the one-body density matrix, or the pair correlation functions. Altogether, given the typical uncertainties affecting the experimental determination of, e.g., the kinetic energy, there scarcely seem to be compelling reasons to pick any of the various refinements of the Aziz I potential (the only likely outcome being that of rendering more complicated the comparison between different theoretical calculations).

As mentioned above, the inclusion of three-body terms in the potential energy of interaction among helium atoms, becomes necessary in order to bring the computed equation of state in agreement with experiment, at pressures significantly higher than those considered in this work (e.g., in the MPa range). On the other hand, the effect on the kinetic energy is rather small \cite{Boninsegni1993,Chang2001}.
\subsubsection{Comparison with recent calculations}
Also shown in Table \ref{tab1} are the most recent estimates \cite{Rugeles2017} published from solid $^4$He in the same range of density considered here, all based on the Aziz III potential. These results were obtained by means of Diffusion Monte Carlo simulations, i.e., they are strictly speaking ground state estimates. However, given the weak dependence on the temperature displayed by our results (shown in Fig. \ref{kinetic}) below 1 K, the comparison of our results, which are at finite temperature, to those of Ref. \cite{Rugeles2017} is quite appropriate. Within DMC,  Bose statistics can be incorporated by a proper choice of trial wave function \cite{Boninsegni2001b}, out of which the exact ground state is (in principle) projected.
\\ \indent
Taken individually (i.e., considering just one specific density) the energy estimates obtained in this work may be considered in reasonable agreement with the DMC ones, in that the differences may not be necessarily regarded as statistically significant. However, the comparison across all density values consistently shows estimates at finite ($T=1$ K) temperature to be {slightly} {    lower} than the (supposedly ``exact'') $T=0$ results. This situation has been observed repeatedly during the past decades, for various Bose systems \cite{Boninsegni2012c,Boninsegni2013,Boninsegni2016}. One may go back to the early GFMC \cite{note3} studies of solid $^4$He based on the Aziz I potential \cite{Kalos1981}, yielding a value of the ground state per $^4$He atom of $-5.175$ K in the solid phase at $n=0.02934$ \AA$^{-3}$, almost 1 K higher than that obtained here (see Table \ref{tab1}).
For a long time, those estimates were held as reliable, largely due to the absence of experimental data and of any alternative calculation.
\\ \indent
Given that the DMC calculations carried out in Ref. \cite{Rugeles2017} have comparable numbers of atoms to those carried out here, and in any case finite-size effects on the energy are relatively small,  the most  plausible explanation for the (small) discrepancy between their results and ours can be ascribed to intrinsic limitations of the DMC projection method.  Fundamental reasons have been proposed as to why finite temperature methods are a superior option to ground state ones, when it comes to studying the ground state of Bose systems \cite{Ceperley1995}, due to various sources of bias from which finite temperature techniques are unaffected. Examples of such sources of bias are the need for a trial wave function, which affects the estimation of {    all} the quantities, as well as the finite population of random walkers, whose effect can be rather large \cite{Boninsegni2001,Boninsegni2012c}.
\\ \indent
There are also small differences between the results for the kinetic energy per particle obtained in this work and those reported in Ref. \cite{Rugeles2017}. This is less noteworthy than the disagreement between total energies yielded by the two calculations, however, first of all because the statistical errors on the kinetic energy quoted in Ref. \cite{Rugeles2017} are fairly large, secondly because it is  well-known that DMC does not allow for unbiased, reliable estimations of thermodynamic averages of observables that do not commute with the energy.

\section{Discussion}\label{concl}

In this paper we present results of extensive numerical studies based on the continuous-space WA, of the low-temperature properties of the solid phase of $^4$He. We focused on a pressure interval ranging from approximately 25 bars (near the $T=0$ melting line) all the way to approximately 140 bars. The calculations made use of several versions of the most popular interparticle pair potential traditionally utilized in computer simulation studies of solid helium. We have compared the results of our simulations to available experimental data, mainly for the EOS and for the kinetic energy per atom. The agreement between theoretical results and existing experimental estimates is satisfactory; within the experimental uncertainties, it is not possible to identify a specific version of the Aziz pair potential that affords a closer reproduction of available experimental data. In general, however, a theoretical model only including pairwise interactions appears sufficient to obtain a reliable quantitative account of the energetics, structure, and dynamics of the helium crystal, at least at moderate pressures.





\funding{This research was funded by the Natural Science and Engineering Research Council of Canada.}

\dataavailability{The computer codes utilized to obtain the results can be obtained by contacting the author.} 

\conflictsofinterest{The author declares no conflict of interest.}


\abbreviations{Abbreviations}{
The following abbreviations are used in this manuscript:\\

\noindent 
\begin{tabular}{@{}ll}
{    bcc} & {Body-centered cubic}\\
{BEC} & Bose-Einstein Condensation\\
DMC & Diffusion Monte Carlo \\
EOS & Equation of State \\
GFMC & Green Function Monte Carlo\\
{    hcp} & Hexagonal close-packed \\
PIMC & Path Integral Monte Carlo\\
$T$ & Temperature \\
WA & Worm Algorithm 
\end{tabular}
}
\reftitle{References}
\externalbibliography{yes}


\bibliography{references}

\begin{thebibliography}{999}

\bibitem[Wilks(1967)]{Wilks1967}
Wilks, J.
\newblock {\em The properties of liquid and solid helium}; Oxford University
  Press: New York,  1967.

\bibitem[Cuccoli et~al.(1993)Cuccoli, Macchi, Tognetti, and Vaia]{cuccoli1993}
Cuccoli, A.; Macchi, A.; Tognetti, V.; Vaia, R.
\newblock Monte Carlo computations of the quantum kinetic energy of rare-gas
  solids.
\newblock {\em Phys. Rev. B} {\bf 1993}, {\em 47},~14923--14931.
\newblock {\url{https://doi.org/10.1103/PhysRevB.47.14923}}.

\bibitem[Timms et~al.(1996)Timms, Evans, Boninsegni, Ceperley, Mayers, and
  Simmons]{Timms1996}
Timms, D.N.; Evans, A.C.; Boninsegni, M.; Ceperley, D.M.; Mayers, J.; Simmons,
  R.O.
\newblock Direct measurements and path integral Monte Carlo calculations of
  kinetic energies of solid neon.
\newblock {\em Journal of Physics: Condensed Matter} {\bf 1996}, {\em 8},~6665.
\newblock {\url{https://doi.org/10.1088/0953-8984/8/36/018}}.

\bibitem[Dusseault and Boninsegni(2017)]{Dusseault2017}
Dusseault, M.; Boninsegni, M.
\newblock Atomic displacements in quantum crystals.
\newblock {\em Phys. Rev. B} {\bf 2017}, {\em 95},~104518.
\newblock {\url{https://doi.org/10.1103/PhysRevB.95.104518}}.

\bibitem[Boninsegni et~al.(2012)Boninsegni, Pollet, Prokof'ev, and
  Svistunov]{Boninsegni2012a}
Boninsegni, M.; Pollet, L.; Prokof'ev, N.; Svistunov, B.
\newblock Role of Bose Statistics in Crystallization and Quantum Jamming.
\newblock {\em Phys. Rev. Lett.} {\bf 2012}, {\em 109},~025302.
\newblock {\url{https://doi.org/10.1103/PhysRevLett.109.025302}}.

\bibitem[Pederiva and Chester(1998)]{Pederiva1998}
Pederiva, F.; Chester, G.V.
\newblock Does antisymmetry matter in b.c.c. He-3 crystals?
\newblock {\em J. Low Temp. Phys.} {\bf 1998}, {\em 113},~741--750.
\newblock {\url{https://doi.org/10.1023/A:1022517914715}}.

\bibitem[Fain~Jr and Lazarus(1970)]{Fain1970}
Fain~Jr, S.C.; Lazarus, D.
\newblock X‐Ray Investigation of Solid Helium.
\newblock {\em J. Appl. Phys.} {\bf 1970}, {\em 41},~1451–1454.
\newblock {\url{https://doi.org/10.1063/1.1659055}}.

\bibitem[Heald et~al.(1983)Heald, Baer, and Simmons]{Heald1983}
Heald, S.; Baer, D.; Simmons, R.
\newblock X-ray diffraction study of thermal vacancies in solid helium-3.
\newblock {\em Solid State Commun.} {\bf 1983}, {\em 47},~807--810.
\newblock {\url{https://doi.org/10.1016/0038-1098(83)90071-6}}.

\bibitem[Mao et~al.(1988)Mao, Hemley, Wu, Jephcoat, Finger, Zha, and
  Bassett]{Mao1988}
Mao, H.K.; Hemley, R.J.; Wu, Y.; Jephcoat, A.P.; Finger, L.W.; Zha, C.S.;
  Bassett, W.A.
\newblock High-Pressure Phase Diagram and Equation of State of Solid Helium
  from Single-Crystal X-Ray Diffraction to 23.3 GPa.
\newblock {\em Phys. Rev. Lett.} {\bf 1988}, {\em 60},~2649--2652.
\newblock {\url{https://doi.org/10.1103/PhysRevLett.60.2649}}.

\bibitem[Loubeyre et~al.(1993)Loubeyre, LeToullec, Pinceaux, Mao, Hu, and
  Hemley]{Loubeyre1993}
Loubeyre, P.; LeToullec, R.; Pinceaux, J.P.; Mao, H.K.; Hu, J.; Hemley, R.J.
\newblock Equation of state and phase diagram of solid $^{4}\mathrm{He}$ from
  single-crystal x-ray diffraction over a large P-T domain.
\newblock {\em Phys. Rev. Lett.} {\bf 1993}, {\em 71},~2272--2275.
\newblock {\url{https://doi.org/10.1103/PhysRevLett.71.2272}}.

\bibitem[Henshaw(1958)]{Henshaw1958}
Henshaw, D.G.
\newblock Structure of Solid Helium by Neutron Diffraction.
\newblock {\em Phys. Rev.} {\bf 1958}, {\em 109},~328--330.
\newblock {\url{https://doi.org/10.1103/PhysRev.109.328}}.

\bibitem[Werthamer(1970)]{Werthamer1970}
Werthamer, N.R.
\newblock Neutron Scattering from Phonons in Solid Helium.
\newblock {\em Phys. Rev. A} {\bf 1970}, {\em 2},~2050--2060.
\newblock {\url{https://doi.org/10.1103/PhysRevA.2.2050}}.

\bibitem[Svensson and Sears(1986)]{Svensson1986}
Svensson, E.; Sears, V.
\newblock Neutron scattering by $^4${He} and $^3${He}.
\newblock {\em Physica B+C} {\bf 1986}, {\em 137},~126--140.
\newblock {\url{https://doi.org/https://doi.org/10.1016/0378-4363(86)90317-7}}.

\bibitem[Simmons(1987)]{Simmons1987}
Simmons, R.O.
\newblock Single-particle dynamics of the solid heliums from deep inelastic
  neutron scattering.
\newblock {\em Can. J. Phys.} {\bf 1987}, {\em 65},~1401--1408.
\newblock {\url{https://doi.org/10.1139/p87-220}}.

\bibitem[Diallo et~al.(2004)Diallo, Pearce, Azuah, and Glyde]{Diallo2004}
Diallo, S.O.; Pearce, J.V.; Azuah, R.T.; Glyde, H.R.
\newblock Quantum Momentum Distribution and Kinetic Energy in Solid
  $^{4}\mathrm{H}\mathrm{e}$.
\newblock {\em Phys. Rev. Lett.} {\bf 2004}, {\em 93},~075301.
\newblock {\url{https://doi.org/10.1103/PhysRevLett.93.075301}}.

\bibitem[Adams et~al.(2007)Adams, Mayers, Kirichek, and Down]{Adams2007}
Adams, M.A.; Mayers, J.; Kirichek, O.; Down, R.B.E.
\newblock Measurement of the Kinetic Energy and Lattice Constant in hcp Solid
  Helium at Temperatures 0.07--0.4 K.
\newblock {\em Phys. Rev. Lett.} {\bf 2007}, {\em 98},~085301.
\newblock {\url{https://doi.org/10.1103/PhysRevLett.98.085301}}.

\bibitem[Diallo et~al.(2007)Diallo, Pearce, Azuah, Kirichek, Taylor, and
  Glyde]{Diallo2007}
Diallo, S.O.; Pearce, J.V.; Azuah, R.T.; Kirichek, O.; Taylor, J.W.; Glyde,
  H.R.
\newblock Bose-Einstein Condensation in Solid $^{4}\mathrm{He}$.
\newblock {\em Phys. Rev. Lett.} {\bf 2007}, {\em 98},~205301.
\newblock {\url{https://doi.org/10.1103/PhysRevLett.98.205301}}.

\bibitem[Azuah et~al.(1997)Azuah, Stirling, Glyde, Boninsegni, Sokol, and
  Bennington]{Azuah1997}
Azuah, R.T.; Stirling, W.G.; Glyde, H.R.; Boninsegni, M.; Sokol, P.E.;
  Bennington, S.M.
\newblock Condensate and final-state effects in superfluid ${}^{4}\mathrm{He}$.
\newblock {\em Phys. Rev. B} {\bf 1997}, {\em 56},~14620--14630.
\newblock {\url{https://doi.org/10.1103/PhysRevB.56.14620}}.

\bibitem[Ceperley(1995)]{Ceperley1995}
Ceperley, D.M.
\newblock Path integrals in the theory of condensed helium.
\newblock {\em Rev. Mod. Phys.} {\bf 1995}, {\em 67},~279--355.
\newblock {\url{https://doi.org/10.1103/RevModPhys.67.279}}.

\bibitem[Herrero and Ramírez(2014)]{Herrero2014}
Herrero, C.P.; Ramírez, R.
\newblock Path-integral simulation of solids.
\newblock {\em J. Phys.: CM} {\bf 2014}, {\em 26},~233201.
\newblock {\url{https://doi.org/10.1088/0953-8984/26/23/233201}}.

\bibitem[Mason and Rice(1954)]{Mason1954}
Mason, E.A.; Rice, W.E.
\newblock {The Intermolecular Potentials of Helium and Hydrogen}.
\newblock {\em J. Chem. Phys.} {\bf 1954}, {\em 22},~522--535.
\newblock {\url{https://doi.org/10.1063/1.1740100}}.

\bibitem[Aziz et~al.(2008)Aziz, Nain, Carley, Taylor, and McConville]{Aziz1979}
Aziz, R.A.; Nain, V.P.S.; Carley, J.S.; Taylor, W.L.; McConville, G.T.
\newblock {An accurate intermolecular potential for helium}.
\newblock {\em J. Chem. Phys.} {\bf 2008}, {\em 70},~4330--4342,
  \href{http://xxx.lanl.gov/abs/https://pubs.aip.org/aip/jcp/article-pdf/70/9/4330/11220077/4330\_1\_online.pdf}{{\normalfont
  [https://pubs.aip.org/aip/jcp/article-pdf/70/9/4330/11220077/4330\_1\_online.pdf]}}.
\newblock {\url{https://doi.org/10.1063/1.438007}}.

\bibitem[Ceperley and Partridge(1986)]{Ceperley1986}
Ceperley, D.M.; Partridge, H.
\newblock {The He$_2$ potential at small distances}.
\newblock {\em J. Chem. Phys.} {\bf 1986}, {\em 84},~820--821.
\newblock {\url{https://doi.org/10.1063/1.450581}}.

\bibitem[Przybytek et~al.(2017)Przybytek, Cencek, Jeziorski, and
  Szalewicz]{Przybytek2017}
Przybytek, M.; Cencek, W.; Jeziorski, B.; Szalewicz, K.
\newblock Pair Potential with Submillikelvin Uncertainties and Nonadiabatic
  Treatment of the Halo State of the Helium Dimer.
\newblock {\em Phys. Rev. Lett.} {\bf 2017}, {\em 119},~123401.
\newblock {\url{https://doi.org/10.1103/PhysRevLett.119.123401}}.

\bibitem[Ceperley and Pollock(1986)]{Ceperley1986b}
Ceperley, D.M.; Pollock, E.L.
\newblock Path-integral computation of the low-temperature properties of liquid
  $^{4}\mathrm{He}$.
\newblock {\em Phys. Rev. Lett.} {\bf 1986}, {\em 56},~351--354.
\newblock {\url{https://doi.org/10.1103/PhysRevLett.56.351}}.

\bibitem[Boninsegni et~al.(2006)Boninsegni, Prokof'ev, and
  Svistunov]{Boninsegni2006}
Boninsegni, M.; Prokof'ev, N.V.; Svistunov, B.V.
\newblock Worm algorithm and diagrammatic Monte Carlo: A new approach to
  continuous-space path integral Monte Carlo simulations.
\newblock {\em Phys. Rev. E} {\bf 2006}, {\em 74},~036701.
\newblock {\url{https://doi.org/10.1103/PhysRevE.74.036701}}.

\bibitem[Prisk et~al.(2017)Prisk, Bryan, Sokol, Granroth, Moroni, and
  Boninsegni]{Prisk2017}
Prisk, T.R.; Bryan, M.S.; Sokol, P.E.; Granroth, G.E.; Moroni, S.; Boninsegni,
  M.
\newblock The Momentum Distribution of Liquid $^4${He}.
\newblock {\em J. Low. Temp. Phys.} {\bf 2017}, {\em 189},~158–184.
\newblock {\url{https://doi.org/10.1007/s10909-017-1798-7}}.

\bibitem[Donnelly and Barenghi(1998)]{Barenghi1998}
Donnelly, R.J.; Barenghi, C.F.
\newblock {The Observed Properties of Liquid Helium at the Saturated Vapor
  Pressure}.
\newblock {\em Journal of Physical and Chemical Reference Data} {\bf 1998},
  {\em 27},~1217--1274.
\newblock {\url{https://doi.org/10.1063/1.556028}}.

\bibitem[Kim and Chan(2004)]{kim2004}
Kim, E.; Chan, M.H.W.
\newblock Observation of Superflow in Solid Helium.
\newblock {\em Science} {\bf 2004}, {\em 305},~1941--1944.
\newblock {\url{https://doi.org/10.1126/science.1101501}}.

\bibitem[Aziz et~al.(1987)Aziz, McCourt, and Wong]{Aziz1987}
Aziz, R.A.; McCourt, F.R.; Wong, C.C.
\newblock A new determination of the ground state interatomic potential for
  He2.
\newblock {\em Mol. Phys.} {\bf 1987}, {\em 61},~1487--1511.
\newblock {\url{https://doi.org/10.1080/00268978700101941}}.

\bibitem[Montgomery et~al.(1989)Montgomery, Petersson, and
  Matsunaga]{Montgomery1989}
Montgomery, J.A.; Petersson, G.A.; Matsunaga, N.
\newblock On the helium pair potential.
\newblock {\em Chem. Phys. Lett.} {\bf 1989}, {\em 155},~413--418.
\newblock {\url{https://doi.org/https://doi.org/10.1016/0009-2614(89)87178-7}}.

\bibitem[Aziz et~al.(1992)Aziz, Slaman, Koide, Allnatt, and Meath]{Aziz1992}
Aziz, R.A.; Slaman, M.J.; Koide, A.; Allnatt, A.R.; Meath, W.J.
\newblock Exchange-coulomb potential energy curves for He-He, and related
  physical properties.
\newblock {\em Mol. Phys.} {\bf 1992}, {\em 77},~321--337.
\newblock {\url{https://doi.org/10.1080/00268979200102471}}.

\bibitem[Aziz et~al.(1995)Aziz, Janzen, and Moldover]{Aziz1995}
Aziz, R.A.; Janzen, A.R.; Moldover, M.R.
\newblock Ab Initio Calculations for Helium: A Standard for Transport Property
  Measurements.
\newblock {\em Phys. Rev. Lett.} {\bf 1995}, {\em 74},~1586--1589.
\newblock {\url{https://doi.org/10.1103/PhysRevLett.74.1586}}.

\bibitem[Janzen and Aziz(1997)]{Janzen1997}
Janzen, A.R.; Aziz, R.A.
\newblock {An accurate potential energy curve for helium based on ab initio
  calculations}.
\newblock {\em J. Chem. Phys.} {\bf 1997}, {\em 107},~914--919.
\newblock {\url{https://doi.org/10.1063/1.474444}}.

\bibitem[Korona et~al.(1997)Korona, Williams, Bukowski, Jeziorski, and
  Szalewicz]{Korona1997}
Korona, T.; Williams, H.L.; Bukowski, R.; Jeziorski, B.; Szalewicz, K.
\newblock {Helium dimer potential from symmetry-adapted perturbation theory
  calculations using large Gaussian geminal and orbital basis sets}.
\newblock {\em J. Chem. Phys.} {\bf 1997}, {\em 106},~5109--5122.
\newblock {\url{https://doi.org/10.1063/1.473556}}.

\bibitem[Przybytek et~al.(2010)Przybytek, Cencek, Komasa, \L{}ach, Jeziorski,
  and Szalewicz]{Przybytek2010}
Przybytek, M.; Cencek, W.; Komasa, J.; \L{}ach, G.; Jeziorski, B.; Szalewicz,
  K.
\newblock Relativistic and Quantum Electrodynamics Effects in the Helium Pair
  Potential.
\newblock {\em Phys. Rev. Lett.} {\bf 2010}, {\em 104},~183003.
\newblock {\url{https://doi.org/10.1103/PhysRevLett.104.183003}}.

\bibitem[Boninsegni et~al.(1994)Boninsegni, Pierleoni, and
  Ceperley]{Boninsegni1993}
Boninsegni, M.; Pierleoni, C.; Ceperley, D.M.
\newblock Isotopic shift of helium melting pressure: Path integral Monte Carlo
  study.
\newblock {\em Phys. Rev. Lett.} {\bf 1994}, {\em 72},~1854--1857.
\newblock {\url{https://doi.org/10.1103/PhysRevLett.72.1854}}.

\bibitem[Chang and Boninsegni(2001)]{Chang2001}
Chang, S.Y.; Boninsegni, M.
\newblock {Ab initio potentials and the equation of state of condensed helium
  at high pressure}.
\newblock {\em J. Chem. Phys.} {\bf 2001}, {\em 115},~2629--2633.
\newblock {\url{https://doi.org/10.1063/1.1386657}}.

\bibitem[Morales et~al.(2009)Morales, Schwegler, Ceperley, Pierleoni, Hamel,
  and Caspersen]{Morales2009}
Morales, M.A.; Schwegler, E.; Ceperley, D.; Pierleoni, C.; Hamel, S.;
  Caspersen, K.
\newblock Phase separation in hydrogen–helium mixtures at Mbar pressures.
\newblock {\em Proc. Natl. Acad. Sci.} {\bf 2009}, {\em 106},~1324--1329,
  \href{http://xxx.lanl.gov/abs/https://www.pnas.org/doi/pdf/10.1073/pnas.0812581106}{{\normalfont
  [https://www.pnas.org/doi/pdf/10.1073/pnas.0812581106]}}.
\newblock {\url{https://doi.org/10.1073/pnas.0812581106}}.

\bibitem[Young et~al.(1981)Young, McMahan, and Ross]{Young1981}
Young, D.A.; McMahan, A.K.; Ross, M.
\newblock Equation of state and melting curve of helium to very high pressure.
\newblock {\em Phys. Rev. B} {\bf 1981}, {\em 24},~5119--5127.
\newblock {\url{https://doi.org/10.1103/PhysRevB.24.5119}}.

\bibitem[Cencek et~al.(2007)Cencek, Jeziorska, Akin-Ojo, and
  Szalewicz]{Cencek2007}
Cencek, W.; Jeziorska, M.; Akin-Ojo, O.; Szalewicz, K.
\newblock Three-Body Contribution to the Helium Interaction Potential.
\newblock {\em J. Phys. Chem. A} {\bf 2007}, {\em 111},~11311--11319.
\newblock {\url{https://doi.org/10.1021/jp072106n}}.

\bibitem[Sesé(2020)]{Sese2020}
Sesé, L.M.
\newblock Real Space Triplets in Quantum Condensed Matter: Numerical
  Experiments Using Path Integrals, Closures, and Hard Spheres.
\newblock {\em Entropy} {\bf 2020}, {\em 22}.
\newblock {\url{https://doi.org/10.3390/e22121338}}.

\bibitem[Mezzacapo and Boninsegni(2006)]{Mezzacapo2006}
Mezzacapo, F.; Boninsegni, M.
\newblock Superfluidity and Quantum Melting of p-H$_2$ Clusters.
\newblock {\em Phys. Rev. Lett.} {\bf 2006}, {\em 97},~045301.
\newblock {\url{https://doi.org/10.1103/PhysRevLett.97.045301}}.

\bibitem[Mezzacapo and Boninsegni(2007)]{Mezzacapo2007}
Mezzacapo, F.; Boninsegni, M.
\newblock Structure, superfluidity, and quantum melting of hydrogen clusters.
\newblock {\em Phys. Rev. A} {\bf 2007}, {\em 75},~033201.
\newblock {\url{https://doi.org/10.1103/PhysRevA.75.033201}}.

\bibitem[Feynman and Hibbs(1965)]{Feynman1965}
Feynman, R.; Hibbs, A.
\newblock {\em Quantum Mechanics and Path Integrals}; McGraw-Hill: New York,
  1965.
\newblock Ch. 10.

\bibitem[Clark and Ceperley(2006)]{Clark2006}
Clark, B.K.; Ceperley, D.M.
\newblock Off-Diagonal Long-Range Order in Solid $^{4}\mathrm{He}$.
\newblock {\em Phys. Rev. Lett.} {\bf 2006}, {\em 96},~105302.
\newblock {\url{https://doi.org/10.1103/PhysRevLett.96.105302}}.

\bibitem[Boninsegni et~al.(2006)Boninsegni, Prokof'ev, and
  Svistunov]{Boninsegni2006b}
Boninsegni, M.; Prokof'ev, N.; Svistunov, B.
\newblock Superglass Phase of $^{4}\mathrm{He}$.
\newblock {\em Phys. Rev. Lett.} {\bf 2006}, {\em 96},~105301.
\newblock {\url{https://doi.org/10.1103/PhysRevLett.96.105301}}.

\bibitem[Boninsegni(2005)]{Boninsegni2005}
Boninsegni, M.
\newblock Permutation Sampling in Path Integral Monte Carlo.
\newblock {\em J. Low Temp. Phys.} {\bf 2005}, {\em 141},~27--46.
\newblock {\url{https://doi.org/10.1007/s10909-005-7513-0}}.

\bibitem[Ceperley et~al.(1996)Ceperley, Simmons, and Blasdell]{Ceperley1996}
Ceperley, D.M.; Simmons, R.O.; Blasdell, R.C.
\newblock Kinetic Energy of Liquid and Solid ${}^{4}\mathrm{He}$.
\newblock {\em Phys. Rev. Lett.} {\bf 1996}, {\em 77},~115--118.
\newblock {\url{https://doi.org/10.1103/PhysRevLett.77.115}}.

\bibitem[Driessen et~al.(1986)Driessen, van~der Poll, and
  Silvera]{Driessen1986}
Driessen, A.; van~der Poll, E.; Silvera, I.F.
\newblock Equation of state of solid $^{4}\mathrm{He}$.
\newblock {\em Phys. Rev. B} {\bf 1986}, {\em 33},~3269--3288.
\newblock {\url{https://doi.org/10.1103/PhysRevB.33.3269}}.

\bibitem[not()]{note1}
The value of the pressure at $n=0.0312$ \AA$^{-3}$ appears to have been
  misquoted in Ref. \cite{Adams2007}.

\bibitem[Boninsegni(2009)]{Boninsegni2009}
Boninsegni, M.
\newblock Quantum statistics and the momentum distribution of liquid
  parahydrogen.
\newblock {\em Phys. Rev. B} {\bf 2009}, {\em 79},~174203.
\newblock {\url{https://doi.org/10.1103/PhysRevB.79.174203}}.

\bibitem[not()]{note2}
The melting line represents the most favorable condition for the possible
  occurrence of off-diagonal long-range order (i.e., BEC). Indeed, what is
  observed is that the decay at large distances of the one body density matrix
  is increasingly more pronounced at higher density.

\bibitem[Boninsegni et~al.(2007)Boninsegni, Kuklov, Pollet, Prokof'ev,
  Svistunov, and Troyer]{Boninsegni2007}
Boninsegni, M.; Kuklov, A.B.; Pollet, L.; Prokof'ev, N.V.; Svistunov, B.V.;
  Troyer, M.
\newblock Luttinger Liquid in the Core of a Screw Dislocation in $^4${He}.
\newblock {\em Phys. Rev. Lett.} {\bf 2007}, {\em 99},~035301.
\newblock {\url{https://doi.org/10.1103/PhysRevLett.99.035301}}.

\bibitem[S\"oyler et~al.(2009)S\"oyler, Kuklov, Pollet, Prokof'ev, and
  Svistunov]{Soyler2009}
S\"oyler, S.G.; Kuklov, A.B.; Pollet, L.; Prokof'ev, N.V.; Svistunov, B.V.
\newblock Underlying Mechanism for the Giant Isochoric Compressibility of Solid
  $^{4}\mathrm{He}$: Superclimb of Dislocations.
\newblock {\em Phys. Rev. Lett.} {\bf 2009}, {\em 103},~175301.
\newblock {\url{https://doi.org/10.1103/PhysRevLett.103.175301}}.

\bibitem[Kuklov et~al.(2022)Kuklov, Pollet, Prokof’ev, and
  Svistunov]{Kuklov2022}
Kuklov, A.B.; Pollet, L.; Prokof’ev, N.V.; Svistunov, B.V.
\newblock Supertransport by Superclimbing Dislocations in $^4${He}.
\newblock {\em Phys. Rev. Lett.} {\bf 2022}, {\em 128},~255301.
\newblock {\url{https://doi.org/10.1103/physrevlett.128.255301}}.

\bibitem[Boninsegni et~al.(2006)Boninsegni, Kuklov, Pollet, Prokof'ev,
  Svistunov, and Troyer]{Boninsegni2006c}
Boninsegni, M.; Kuklov, A.B.; Pollet, L.; Prokof'ev, N.V.; Svistunov, B.V.;
  Troyer, M.
\newblock Fate of Vacancy-Induced Supersolidity in $^{4}\mathrm{He}$.
\newblock {\em Phys. Rev. Lett.} {\bf 2006}, {\em 97},~080401.
\newblock {\url{https://doi.org/10.1103/PhysRevLett.97.080401}}.

\bibitem[Boninsegni and Prokof'ev(2012)]{Boninsegni2012b}
Boninsegni, M.; Prokof'ev, N.V.
\newblock Colloquium: Supersolids: What and where are they?
\newblock {\em Rev. Mod. Phys.} {\bf 2012}, {\em 84},~759--776.
\newblock {\url{https://doi.org/10.1103/RevModPhys.84.759}}.

\bibitem[Boninsegni(2012)]{Boninsegni2012}
Boninsegni, M.
\newblock Supersolid Phases of Cold Atom Assemblies.
\newblock {\em J. Low Temp. Phys.} {\bf 2012}, {\em 168},~137--149.
\newblock {\url{https://doi.org/10.1007/s10909-012-0571-1}}.

\bibitem[Rugeles et~al.(2017)Rugeles, Ujevic, and Vitiello]{Rugeles2017}
Rugeles, E.J.; Ujevic, S.; Vitiello, S.A.
\newblock Solid $^{4}\mathrm{He}$ and the diffusion Monte Carlo method: A study
  of their properties.
\newblock {\em Phys. Rev. E} {\bf 2017}, {\em 96},~043306.
\newblock {\url{https://doi.org/10.1103/PhysRevE.96.043306}}.

\bibitem[Boninsegni et~al.(2001)Boninsegni, Lee, and Crespi]{Boninsegni2001b}
Boninsegni, M.; Lee, S.Y.; Crespi, V.H.
\newblock Helium in One-Dimensional Nanopores: Free Dispersion, Localization,
  and Commensurate/Incommensurate Transitions with Nonrigid Orbitals.
\newblock {\em Phys. Rev. Lett.} {\bf 2001}, {\em 86},~3360--3363.
\newblock {\url{https://doi.org/10.1103/PhysRevLett.86.3360}}.

\bibitem[Boninsegni and Moroni(2012)]{Boninsegni2012c}
Boninsegni, M.; Moroni, S.
\newblock Population size bias in diffusion Monte Carlo.
\newblock {\em Phys. Rev. E} {\bf 2012}, {\em 86},~056712.
\newblock {\url{https://doi.org/10.1103/PhysRevE.86.056712}}.

\bibitem[Boninsegni(2013)]{Boninsegni2013}
Boninsegni, M.
\newblock Ground State Phase Diagram of Parahydrogen in One Dimension.
\newblock {\em Phys. Rev. Lett.} {\bf 2013}, {\em 111},~235303.
\newblock {\url{https://doi.org/10.1103/PhysRevLett.111.235303}}.

\bibitem[Boninsegni(2016)]{Boninsegni2016}
Boninsegni, M.
\newblock Absence of superfluidity in a parahydrogen film intercalated within a
  crystal of Na atoms.
\newblock {\em Phys. Rev. B} {\bf 2016}, {\em 93},~054507.
\newblock {\url{https://doi.org/10.1103/PhysRevB.93.054507}}.

\bibitem[not()]{note3}
GFMC is a methodology closely related to DMC, in many respects a precursor
  thereof.

\bibitem[Kalos et~al.(1981)Kalos, Lee, and Whitlock]{Kalos1981}
Kalos, M.H.; Lee, M.A.; Whitlock, P.A.
\newblock Modern potentials and the properties of condensed Helium-four.
\newblock {\em Phys. Rev. B} {\bf 1981}, {\em 24},~115--130.

\bibitem[Boninsegni(2001)]{Boninsegni2001}
Boninsegni, M.
\newblock Phase Separation in Mixtures of Hard Core Bosons.
\newblock {\em Phys. Rev. Lett.} {\bf 2001}, {\em 87},~087201.
\newblock {\url{https://doi.org/10.1103/PhysRevLett.87.087201}}.

\end{thebibliography}

%


\end{document}